\documentclass[useAMS,usenatbib,usegraphicx,onecolumn]{mn2e}

\title[Invariant manifolds and spiral arms in galaxies]
       {The coalescence of invariant manifolds and the spiral
        structure of barred galaxies}

\author[P. Tsoutsis, C. Efthymiopoulos, and N. Voglis]
       {
       P. Tsoutsis$^{1,2}$, C. Efthymiopoulos$^1$, and N. Voglis$^\dag$\\
       $^1$Research Center for Astronomy and Applied Mathematics,
           Academy of Athens, Soranou Efessiou 4, GR-115 27 Athens, Greece\\
       $^2$Section of Astronomy, Astrophysics and Mechanics, Department of
           Physics, University of Athens,\\
           Panepistimiopolis, GR-157 84 Zografos, Athens, Greece\\
           e-mail: ptsoutsi@phys.uoa.gr, cefthim@academyofathens.gr
       }

\date{Released 2008 Xxxxx XX}
\pagerange{\pageref{firstpage}--\pageref{lastpage}} \pubyear{2008}

\def\LaTeX{L\kern-.36em\raise.3ex\hbox{a}\kern-.15em
    T\kern-.1667em\lower.7ex\hbox{E}\kern-.125emX}

\begin{document}
\label{firstpage}
\maketitle

\begin{abstract}
In a previous paper (Voglis et al. 2006a, paper I) we demonstrated
that, in a rotating galaxy with a strong bar, the unstable
asymptotic manifolds of the short period family of unstable periodic
orbits around the Lagrangian points L$_1$ or L$_2$ create
correlations among the apocentric positions of many chaotic orbits,
thus supporting a {\it spiral} structure beyond the bar. In the
present paper we present evidence that the unstable manifolds of
{\it all} the families of unstable periodic orbits near and beyond
corotation contribute to the same phenomenon. Our results refer to a
N-Body simulation, a number of drawbacks of which, as well as the
reasons why these do not significantly affect the main results, are
discussed. We explain the dynamical importance of the invariant
manifolds as due to the fact that they produce a phenomenon of
`stickiness' slowing down the rate of chaotic escape in an otherwise
non-compact region of the phase space. We find a stickiness time of
order $100$ dynamical periods, which is sufficient to support a
long-living spiral structure. Manifolds of different families become
important at different ranges of values of the Jacobi constant. The
projections of the manifolds of all the different families in the
configuration space produce a pattern due to the `coalescence' of
the invariant manifolds. This follows closely the maxima of the
observed $m=2$ component near and beyond corotation. Thus, the
manifolds support both the outer edge of the bar and the spiral
arms.
\end{abstract}


\section{Introduction}
The dynamics of spiral arms in rotating galaxies with a strong
non-axisymmentric perturbation is probably very different from that
in galaxies with a weak non-axisymmentric perturbation (see
Contopoulos 2004 for a review). In the case of weak perturbation,
bars or oval distortions have small, if not zero amplitude, and the
main constituents of the non-axisymmetric component of the matter
are the spiral arms themselves. The spirals are successfully
described by models based on {\it stable} periodic orbits and the
quasi-periodic orbits around them. Chaos is of little or no
importance in such models. In the case of strong perturbation, on
the other hand, the spirals start near the ends of strong bars, and
they may extend to large distances beyond corotation. Furthermore,
many orbits near and beyond corotation are chaotic, and some of them
(the so called `hot population' (Sparke and Sellwood 1987)) can
wander both inside and outside corotation. In self-consistent models
of barred spiral galaxies it has been indicated for the first time
by Kaufmann and Contopoulos (1996) that these chaotic orbits play a
significant role by supporting both the bar and the spiral arms.
More recently, N-body models (Voglis et al. 2006b), or response
models fitting a real barred spiral galaxy (Patsis 2006), were
constructed in which the spiral arms are supported almost entirely
by chaotic orbits.

To construct a theory of spiral structure based on orbits  means essentially to
provide a theoretical mechanism explaining how the angular positions of the apsides
(apocenters) of the orbits become delineated in a way reproducing self-consistently
an observed spiral pattern. In the case of weak perturbation, the alignment is produced
by the stable periodic orbits of the $x_1$ family which form `precessing ellipses', i.e.,
a family of ellipses with apocenters changing their azimuthal position, as a function
of the Jacobi constant, in a direction along the spiral arms (see Grosb{\o}l 1994
for a review). The termination of the main spiral is placed near the 4/1 resonance
(Contopoulos 1985, Contopoulos and Grosb{\o}l 1986, 1988, Patsis et al. 1991, 1994,
Patsis and Kaufmann 1999), but weak extensions may also be found beyond the 4/1
resonance, reaching the corotation region.

However, these models are not applicable in the case of spirals in a galaxy with a
strong bar, in which the orbits supporting the spiral are chaotic. Precisely, in our
previous paper (Voglis et al. 2006a, hereafter paper I) we proposed a mechanism
yielding the alignment of the apocenters of the chaotic orbits that is necessary in
order to produce a spiral pattern. This is based on the {\it invariant manifolds}
of the family of short period unstable periodic orbits which exist around the
unstable equilibria $L_1$ or $L_2$. Besides our study, the role of these manifolds
has been pointed out in the formation both of rings (Romero-Gomez et al. 2006) and
spiral arms (Romero-Gomez et al. 2007).

\begin{figure}
\centering
\includegraphics[scale=1]{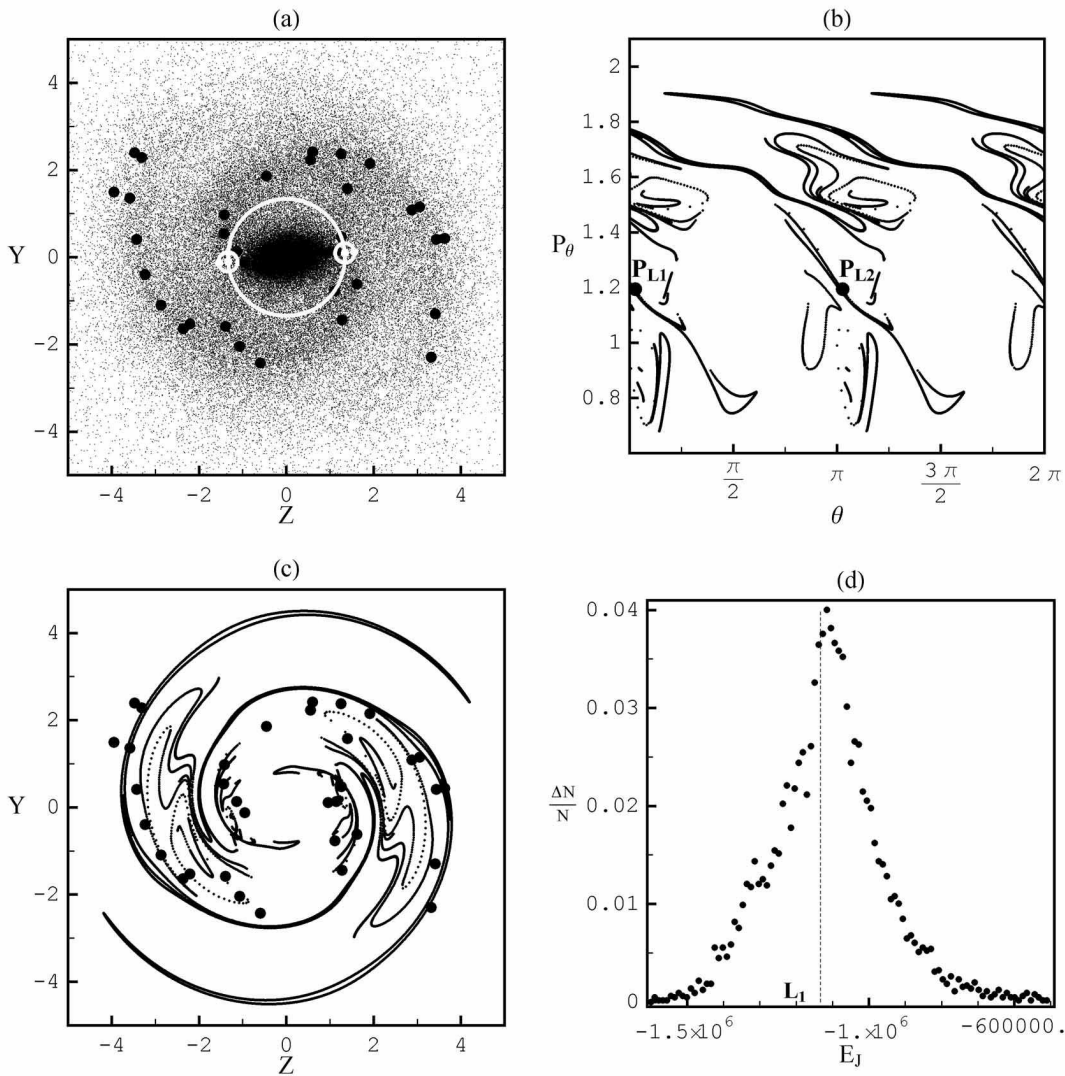}
\caption{(a) A snapshot of the N-Body simulation (same as in Fig.5a
of paper I). The thick dots show local maxima of the projected
surface density. The big white circle passes through the unstable
Lagrangian points $L_1$, $L_2$ and it is refered to as the
corotation circle. The small circles around $L_1$ and $L_2$ are the
short period unstable periodic orbits $PL_1$, $PL_2$. (b) Projection
of the invariant manifolds of the $PL_1$ and $PL_2$ orbits in the
surface of section $(\theta,p_\theta)$, with $p_r=\dot{r}=0$ and
$\dot{p}_r<0$ (condition of the apocenters), for the value of the
Jacobi constant $E_J=-1.125\times 10^6$. (c) Projection of the same
manifolds as in (b) in the configuration space $(y,z)$. The thick
dots are the same as in (a). (d) The distribution of the values of
the Jacobi constants of all the N-body particles which are in small
bins $r\Delta r\Delta\theta$ around the maxima of the surface
density, i.e., the thick dots of (a). The distribution is normalized
according to the total number of particles in all the bins. }
\label{}
\end{figure}

The basic paradigm for this mechanism is summarized in Figure 1. Keeping the same
notation as in paper I, we deal with an N-Body system (Fig.1a) which is flat on
the plane $(y,z)$, called the disk plane, and has also some thickness along the
x-axis, vertical to the disk (details on the numerical runs and features of the
system are given in subsection 2.1 below).  Fig.1a shows the projection of the
particles on the disk plane at the snapshot $t=47$ (after seven pattern rotations).
The thick dots mark the
positions of local maxima of the projected surface density, derived by counting the
particles in successive rings from the center on the plane $(y,z)$. In the snapshot
shown in Fig.1a the system has a strong bar as well as a conspicuous bi-symmetric
spiral pattern which lies almost entirely outside corotation (marked by a white
circle passing through $L_{1,2}$).

The small white circles in Fig.1a around $L_1$ and $L_2$ are orbits belonging to
the family of {\it short period} unstable periodic orbits, which bifurcate from
$L_1$ or $L_2$ and form loops around these points. The period of the loop is of the
same order as the value of the epicyclic period at corotation. The family of these
orbits, which are hereafter called $PL_1$ orbits (or $PL_2$, around $L_2$),
exists for values of the Jacobi constant $E_J>E_{J,L_1}$. Here, as in
paper I, we define the Jacobi constant by a 2D Hamiltonian approximation
on the disk plane, namely
\begin{equation}\label{jacnb}
E_J={1\over 2}(v_y^2+v_z^2)-\Omega_p(zv_y-yv_z) + V(x=0,y,z)~~.
\end{equation}
The potential function $V(x,y,z)$ is found from the N-Body code in a closed
mathematical form expressed as a series of basis functions with coefficients
calculated via a `smooth field' technique (see subsection 2.1 below).

For a fixed value of $E_J$ there is one short period orbit, $PL_1$, around $L_1$
and the symmetric orbit $PL_2$ around $L_2$ (Fig.1a). Let $(r,p_r)$,
$(\theta,p_\theta)$ be canonical pairs of variables in the Hamiltonian
describing the orbits:
\begin{equation}\label{hamgen}
H(r,\theta,p_r,p_{\theta})\equiv{1\over 2}\big(p_r^2+{p_\theta^2\over r^2}\big)
-\Omega_p p_\theta +V(r,\theta)
\end{equation}
where $V(r,\theta)\equiv V(x=0,r\sin\theta,r\cos\theta)$. The space of the
variables $(r,\theta,p_r,p_\theta)$ is called the phase space. Let
$r(t)$, $\theta(t)$, $p_r(t)$, $p_\theta(t)$ denote an orbit with initial
conditions $(r_0,\theta_0,p_{r0},p_{\theta0})$ under the time flow (time$=t$)
of the Hamiltonian (\ref{hamgen}).
The {\it unstable manifold} of $PL_1$ is defined as the set of initial conditions
$(r_0,\theta_0,p_{r0},p_{\theta0})$ in phase space tending asymptotically to $PL_1$
in the backward sense of time, namely
\begin{equation}
{\cal W}^U_{PL_1} = \big\{\bigcup(r_0,\theta_0,p_{r0},p_{\theta0}):
\lim_{t\rightarrow -\infty}||Q(t;r_0,\theta_0,p_{r0},p_{\theta0})-PL_1||=0
\big\}
\end{equation}\label{wu}
where $Q(t;r_0,\theta_0,p_{r0},p_{\theta0})$ denotes the position of
a particle in an orbit starting with the above initial conditions,
and the norm $||\cdot||$ is defined as the minimum of the distances
of $Q(t)$ from the locus of all the points in phase space belonging
to the orbit $PL_1$. In the forward sense of time, the orbits with
initial conditions taken on ${\cal W}^U_{PL_1}$ and near $PL_1$
deviate exponentially from $PL_1$ with a rate determined by the
largest eigenvalue of the linearized flow around $PL_1$. However,
the linearized approximation is not valid at large distances from
$PL_1$ and the invariant manifold ${\cal W}^U_{PL_1}$ acquires in
general a quite complicated form. In particular, the intersections
of the unstable manifold ${\cal W}^U_{PL_1}$ with the {\it stable
manifold}
\begin{equation}\label{ws}
{\cal W}^S_{PL_1} = \big\{\bigcup(r_0,\theta_0,p_{r0},p_{\theta0}):
\lim_{t\rightarrow \infty}||Q(t;r_0,\theta_0,p_{r0},p_{\theta0})-PL_1||=0
\big\}
\end{equation}
produce the so-called `homoclinic tangle' which is the main source of
chaos locally, in a domain of the phase space surrounding
$PL_1$.

The unstable and stable manifolds ${\cal W}^U_{PL_1}$, ${\cal
W}^S_{PL_1}$ are two-dimensional sets embedded in the
three-dimensional hypersurface of fixed Jacobi constant. Let
$p_r=\dot{r}=0$, $\dot{p}_r<0$ be the plane of the phase space
corresponding to the set of points where orbits reach their
apocenters. The intersection of ${\cal W}^U_{PL_1}$ with the plane
$p_r=\dot{r}=0$ defines a {\it one-dimensional} set that we still
call the unstable manifold of $PL_1$. Fig.1b shows the projection of
the manifolds ${\cal W}^U_{PL_1}$ and ${\cal W}^U_{PL_2}$ in the
plane $(\theta,p_\theta)$. The plane $(\theta,p_\theta)$ is
hereafter called the surface of section. We clearly observe the
intricate shape of the unstable invariant manifold that results in a
strongly chaotic behavior of the orbits near $PL_1$ and $PL_2$.
Fig.1c shows the projection of the {\it same} manifolds on the plane
$(y=r\sin\theta, z=r\cos\theta)$, called the configuration space,
which is the usual space where the motion of stars takes place.
Although the projection of the invariant manifolds on the plane
$(y,z)$ is rather intricate, it is clear that most of the time the
manifolds follow closely the spiral pattern that is defined by the
maxima of the particles' surface density (a detailed comparison of
the two patterns is made in section 3 below). In fact, every time
when a chaotic orbit with initial conditions on these manifolds
reaches an apocentric position, this position is always at another
point of the same mamifolds. Thus, the apocenters of all the chaotic
orbits of the manifold create an invariant locus as shown in Fig.1c.
Hence, in paper I we supported that such manifolds yield, precisely,
the necessary mechanism of alignment of the apocenters that induces
the observed spiral pattern of the system.

Now, the manifolds shown in Figs.1b,c correspond to a particular
value of the Jacobi constant $E_J=-1.125\times 10^6$ in the N-Body
units introduced in paper I (summarized in subsection 2.1 below).
However, the particles forming the spiral are spread in a wide range
of values of the Jacobi constant. Figure 1d shows the distribution
of Jacobi constants for the particles counted within small cells
$r\Delta r\Delta\theta$, $\Delta r=0.2$, $\Delta\theta= \pi/26$,
centered at the thick dots of Fig.1a, namely, along the spiral. The
value $E_J=-1.125\times 10^6$ is very close to the peak of the
distribution, but the distribution has a wide dispersion, of order
$\sigma=5\times 10^5$. The one $\sigma$ level of the distribution is
in the range $E_{J,min}\simeq -1.25\times 10^6$, $E_{J,max}\simeq
-0.95\times 10^6$. This means that the manifolds shown in paper I
correspond to a very narrow band of energies compared to the real
distribution. Since many families of unstable periodic orbits
coexist at different values of the Jacobi constant, the question is
then whether the invariant manifolds of these families play also
some role in the observed spiral pattern.

In the present paper we consider the invariant manifolds of many
different periodic orbits covering a significant part of the range
of Jacobi constants in the histogram of Fig.1d. Let us note that the
$PL_{1,2}$ family, i.e., the only family examined in paper I, does
not exist for values of the Jacobi constant
$E_J<E_{J,L_1}=-1.133\times 10^6$, whereas Fig.1d clearly shows that
about half of the mass in this histogram is in particles with
$E_J<E_{J,L_1}$. Thus, for these particles it is necessary to
consider the invariant manifolds of families other than $PL_1$ in
order to complete the picture of paper I. But even when we
considered the domain $E_J\geq E_{J,L_1}$, permissible to $PL_1$, we
found that the invariant manifolds of other families play a role in
determining the dynamics of chaotic orbits and, thereby, the spiral
structure. In particular, we found that the invariant manifolds of
{\it all} the examined families of unstable periodic orbits produce
essentially the same pattern in configuration space, i.e., the same
spiral pattern. These manifolds produce a phenomenon of
`stickiness', namely while the phase space is, in general, open to
fast chaotic escapes (with escape times of only a few radial
periods), the chaotic orbits with initial conditions on or close to
an invariant manifold can only escape by following this manifold,
and then the escape time increases considerably, i.e., it is of
order $10^2$ periods. During this time the chaotic orbits support
the spiral pattern. In fact, while many different families of
periodic orbits co-exist at one value of the Jacobi constant, we
find that only a subset of them create the dominant stickiness
phenomena at that particular value. As analyzed below, the above
phenomena can be explained on the basis of known properties of the
invariant manifolds in conservative dynamical systems. When applied
to the manifolds of the periodic orbits of the galaxy, these
properties allow us to construct a consistent picture of the
mechanism causing the alignment of the apocenters of the chaotic
orbits along the spiral arms.

The paper is structured as follows: Section 2 describes briefly the simulation
as well as the phase space structure of the system under study at various values
of the Jacobi constant. This is compared to the distribution of the real particles
of the N-body simulation. Section 3 presents the analysis of the invariant manifolds
for nine different families of periodic orbits. Plots are given of the unstable
manifolds of these orbits in phase space as well as in configuration space,
which are compared with the distribution of the N-Body particles in both
spaces. The observed spiral pattern is reconstructed by superposing these plots.
We finally explore which families are dynamically important at various ranges of
values of the Jacobi constant. Section 4 summarizes the main conclusions of the
present study.

\begin{figure}
\centering
\includegraphics[scale=1]{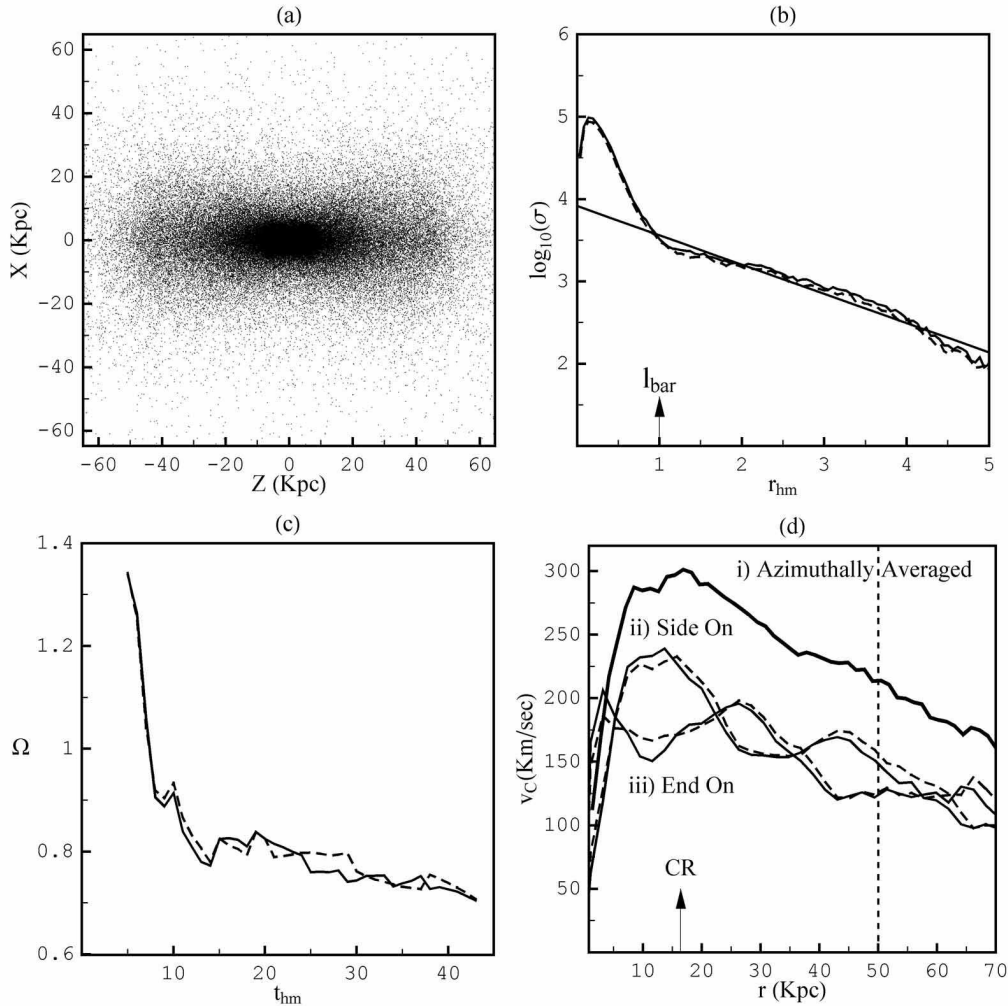}
\caption{(a) The N-body system viewed edge-on (bar nearly line-on).
(b) The projected surface density profile on the disk plane (y-z)
for the experiment with $1.3\times 10^5$ particles (solid line)
versus the same profile for the disk particles of the experiment
with $1.17\times 10^6$ particles (dashed line). The straight solid
line is an exponential fit to the surface density profile for the
radii $r\geq l_{bar}=1$ (see text).(c) Comparison of the time
evolution of the pattern speed, and (d) comparison of the rotation
curves for the two experiments (solid lines = experiment with
$1.3\times 10^5$ particles, dashed lines = experiment with
$1.17\times 10^6$ particles). The top bold solid curve is the
`azimuthally averaged' rotation curve $v_c(r)=<L(r)>/r$ where
$<L(r)>$ is the mean angular momentum perpendicularly to the disc
for the particles in annuli of mean radius $r$ (see text).} \label{}
\end{figure}

\section{Simulation and phase space structure}

\subsection{The N-Body experiment}
The details of the numerical simulation have been presented in
Voglis et al. (2006b). A brief account follows below.

Our analysis refers to a particular simulation called `QR3' in
Voglis et al. (2006b). This experiment starts with a triaxial
equilibrium system which has initially no rotation. This system is
obtained by running a 'collapse' simulation. The particles are
initially taken to have small radial position and velocity
perturbations from an ideal Hubble flow. These perturbations
correspond to a rms radial profile of mass perurbations in the early
Universe of the form $\delta\mu/\mu\propto r^{-(n+3)/2}$, where the
power exponent $n$, derived from the spectrum of density
perturbations at decoupling, has the value $n=-2$ for perturbations
corresponding to the galactic mass scale (fixed as $10^{12}$ solar
masses). The initial positions and velocities of the particles are
determined using the analytical formulae of Palmer and Voglis (1983)
which yield practically equivalent results to the Zel'dovich (1970)
approximation (a detailed description of the methodology of
production of this type of initial conditions is given in sect. 3 of
Efthymiopoulos and Voglis 2001). The numerical simulation starts
well before the bound system has reached its maximum expansion,
before collapsing. The initial redshift depends somewhat on the used
value of the Hubble parameter but it is well before $z=10$. Thus,
the Universe at the starting point of the simulation is to a great
precision Einstein-de Sitter, even if the cosmological constant
$\Lambda$ is non-zero. Furthermore, the system is considered
isolated from its environment, thus the expansion of the Universe
does not affect it any longer, i.e., the system is simulated by
Newtonian forces (accelerations due to $\Lambda$ can be safely
ignored at the galactic scale , e.g. in our galaxy
$g_\Lambda/g_{gravity}\sim 10^{-5}\times (r/8Kpc)^2$). Thus, the
above initial conditions are consistent with both an Einstein-de
Sitter or $\Lambda$CDM model of the Universe with $\Omega=1$.

The violent relaxation phase is simulated using the tree code of Hernquist
(1987), while the subsequent slow approach to equilibrium, through `phase mixing',
is simulated by an improved version of the self-consistent field (SCF) code of Allen
et al. (1990). The equilibrium system is triaxial, and quite elongated, as a result
of the onset of a radial orbit instability (the mean ellipticity on the plane of
long and short axes is E6). Axes are oriented so that $z$ coincides with the
long axis, $y$ with the intermediate axis, and $x$ with the small axis of
the moment of inertia ellipsoid.

The triaxial equilibrium system serves as the basis for the production of
a sequence of rotating systems which preserve the energy and virial equilibrium,
but have different amounts of total angular momentum, i.e., different spin
parameters $\lambda$. The detailed procedure by which rotation is added
is described in section 2 of Voglis et al (2006b). It essentially consists
of re-orienting, at a particular moment, the component $\vec{v}_{yz}$ of
the velocity of each particle on the y-z plane so that the new vector
$\vec{v}_{yz}'$ has equal size as $\vec{v}_{yz}$ and is directed
perpendicularly to the particle's position vector $\vec{r}$ with respect to
the center of mass of the system. By repeatedly implementing the velocity
re-orientation procedure, after a system is allowed to relax for 20 half-mass
crossing times at each step, we obtain a sequence of systems with progressively
higher spin parameter.

The QR3 system is third in this sequence (with spin parameter $\lambda=0.22$).
As in paper I, in the present
paper we present results from a realization of the experiment using $1.3\times
10^5$ particles. However, as demonstrated below the features of the system remain
practically invariant when the simulation is repeated using nine times as many
particles ($1.17\times 10^6$).

We fix units by considering that the system contains a total mass equal to
$M=10^{12}$ solar masses. The unit of length is the half-mass radius of the
relaxed triaxial system at the begining of the simulation. For a Hubble
parameter $H_0=72$Km/sec/Mpc, this turns out to be $R_{hm}=12.6$Kpc in
physical units. The unit time is set equal to the half-mass crossing time
$t_{hm}$ which in physical units is equal to 32Myr. The unit of velocity
is, thus, $V_{unit}=0.48$Km/sec, while the unit of energy (per unit mass)
is taken as $E_{unit}=V_{unit}^2$. Finally, the unit of angular momentum is
taken equal to the value of the angular momentum for a circular orbit at
corotation, $L_{unit}=\Omega_pR_c^2$. In all subsequent plots we use the
above units (except for figs.2a,d in which we use physical units).
The length unit must be rescaled by a factor $M^{1/2}/10^6$ if the mass
is not equal to the adopted value $10^{12}$.

The morphological details of the QR3 experiment are given in section
3 of Voglis et al. (2006b). The main features of the system can be
summarized as follows: The particles have no a priori identities as
of belonging to a disk or a halo. The system is flattened as a
whole, and when viewed edge on it yields an axial ratio $\sim 0.3$
(Fig.2a) (this was called a `thick disk' in Voglis et al. 2006b). On
the other hand, particles forming a thin disk can be identified by
taking any small arbitrary value of the half-thickness $\Delta x$.
We then find that there are particles which remain confined within a
vertical distance $\Delta x$ from the equatorial plane during the
whole simulation up to $t=47$. By varying $\Delta x$ the number of
particles found to satisfy this criterion also varies. Adopting a
conventional ratio of the disk over halo mass in a galaxy
$M_{disk}/M_{halo} = 1:5=0.2$, we vary $\Delta x$ until getting the
value for which the percentage of particles staying confined to
$\pm\Delta x$ becomes equal to $0.2$. Doing so, we found $\Delta x =
0.145R_{hm}\simeq 2.2$Kpc. In the sequel we conventionally identify
these particles as belonging to the disk, while the remaining
particles are identified as halo particles. Such a distinction is a
shortcoming of the simulation, since there is no a priori guarantee
that the particles confined so far in the so-defined `disk' do not
escape from it at subsequent time steps. A better analysis would
necessitate running separately the disk and halo particles, as in
many standard simulations in the literature. One main reason for the
present type of simulation, however, is the possibility to use an
SCF code to produce the potential, so that the analysis of the phase
space structure is possible in terms of smooth orbits.

In the simulation with $1.17\times 10^6$ particles we have about $2.3\times
10^5$ particles in the disk. The surface density profile $\sigma(R)$ of these
particles (Fig.2b) compares very well with the projected surface density
profile of the whole configuration in the simulation with $1.3\times 10^5$
particles. Thus, in both cases the projected surface density shows asymptotically
(for large $r$) an exponential profile $\sigma(r)\propto \exp(-r/R_D)$ with a
scale length $R_D\simeq 0.82R_{hm}=10.3$Kpc. Taking the optical radius as
$R_0\simeq 3R_D\simeq 31$Kpc, the disk thickness ratio is $2.2/31\simeq 0.07$.

The time evolution of the pattern speed $\Omega_p(t)$ is also quite similar
in the simulations with $10^5$ and $10^6$ particles (Fig.2c). There is
initially some angular deceleration which slows down considerably after a time
$t=10t_{hm}$. At the snapshot $t=47$ (corresponding to $t=1.5$Gyr in physical
units) the pattern speed is almost stabilized at a value $\Omega_p=0.67/t_{hm}$
or $\Omega_p=20.6$Km/sec/Kpc. The pattern has performed about seven revolutions
by that time. The length of the major semi-axis of the bar can be estimated from
the inner point (arrow of fig.2b) at which the profile $\sigma(R)$ starts deviating
from exponential. We find $l_{bar}\sim 1R_{hm} = 12.6$Kpc. Corotation (taken
as the average of the distances to $L_1$ and $L_4$ is at $R_c\simeq 1.3\times
l_{bar} = 16.4$Kpc. Thus we have a strong, relatively slow and long bar
which is reminiscent of the bars found in old and strongly barred galaxies
(see e.g. Lindblad et al. 1996 for indicative values of the same parameters
in the case of NGC1365, or Aguerri et al. (1998), Gadotti and de Souza (2006)
and Michel-Dansac and Wozniak (2006) for more general references). In fact,
the above given values for $l_{bar}$ and $R_c$ are somewhat larger than
typical, but they should all be rescaled to smaller values if $M<10^{12}$.

The rotation curves obtained by calculating the mean values of the
line-of-sight velocity profiles of the particles when the system is
viewed edge-on and the bar is either side-on or end-on are shown in
Fig.2d. The bold curve on top of the two other curves represents an
`azimuthally averaged' rotation curve, obtained via
$v_c(r)=<L(r)>/r$, where $<L(r)>$ is the mean component of the
angular momentum perpendicularly to the disk of all the particles in
an annulus $r\pm\Delta r/2$ divided by the central radius of the
annulus $r$. This calculation is the same as in Voglis et al. 2006b,
fig.3e, but now it is given for the snapshot $t=47$. The horizontal
axis is also in the same limits as in that figure. We notice that
the azimuthally averaged rotation curve is declining, and the
decline becomes Keplerian after a radius $R=\sim
3.2R_c=4R_{hm}=50$Kpc. This is due to the fact that the numerical
code imposes a truncation radius for the whole system, i.e. the disk
and the halo, after which Laplace's equation is solved instead of
Poisson's equation. Nevertheless, the decline below this radius is
less steep and the system partly mimics the effects of a disk
embedded in a halo. More precisely, the peak value is at 300Km/sec
and the value at 50Kpc is 220Km/sec. Thus, the decline is 27\%. This
picture is again reminiscent of strongly barred galaxies like
NGC3515 (see the rotation curve in Lindblad et al. 1996). Similar
conclusions are drawn by inspecting the rotation curves obtained by
the line-of-sight velocity profiles. Since such profiles are biased
towards their left wing, the rotational velocities obtained from
then are smaller than those obtained by the azimuthal averaging. The
difference increases as the velocity dispersion increases, and in
our case it is of order 50Km/sec. In particular, the peak value of
the side-on curve is at $v_c=250$Km/sec while the curve stabilizes
at about a 35\% lower value at $R=1.5R_c$ and remains flat up to
about $1.2R_o=40$Kpc, where $R_o=3.2R_D$ is the disk's optical
radius. All our plots below of the configuration space are in square
boxes of dimension $5R_{hm}= 63$Kpc, thus, phenomena near the edge
of these boxes are not so reliable. However, this does not affect in
any essential way the behavior of the invariant manifolds well below
this distance.

Another unpleasant feature of the simulation is the high velocity dispersion of the
disk beyond the bar, with dispersions exceeding $100$Km/sec. This is due to the
fact that the simulation is purely dissipationless, i.e., there are no gas effects
simulated. Thus, both the disk and the halo become 'hot' quickly. While this does
not prevent us from detecting spiral $m=2$ components beyond the bar, the high value
of the velocity dispersion causes the space between the spirals and the bar to
be substabtially populated, while in observed galaxies this space often appears
to be empty.

Finally, the amplitude of the spiral pattern undergoes oscillations
during the whole run. This is an interesting phenomenon which is
described in detail in subsection 2.5 of Voglis et al. 2006b. Up to
$t=50$ the ratio $A_2/A_0$ of the Fourier decomposition of the
surface density profile exhibits variations. The bar maximum is
about $0.8$. On the other hand, the $A_2/A_0$ ratio for the spiral
arms oscillates, roughly between the limits 0.2 and 0.5. The
frequency of this oscillation is resonant with the bar. In fact,
there appear to be two oscillating modes of the spiral pattern. An
inner part, near $L_1$ or $L_2$, rotates almost together with the
bar. An outer part (beyond $r=4$ length units) rotates more slowly
than the bar. As a result, the two parts appear sometimes disjoint,
and at other times they are rejoined. At snapshots when the two
parts are joined smoothly, the whole spiral pattern becomes
conspicuous, i.e., the amplitude $A_2/A_0$ becomes maximal over a
large radial extent. The snapshots $t=18$, $t=47$ and $t=74$,
treated in Paper I, are close to such a maximum. At such snapshots
we typically find that the invariant manifolds trace well the spiral
arms in a considerable part of the manifolds length. Although the
above description renders immediately clear that there is more to
consider in the dynamics of the spiral arms than simply the
invariant manifolds, it is also clear that the latter are a key
ingredient of the dynamics. This is further explored in the sequel.

\subsection{Particle distribution and phase space structure}

\begin{figure}
\centering
\includegraphics[scale=1]{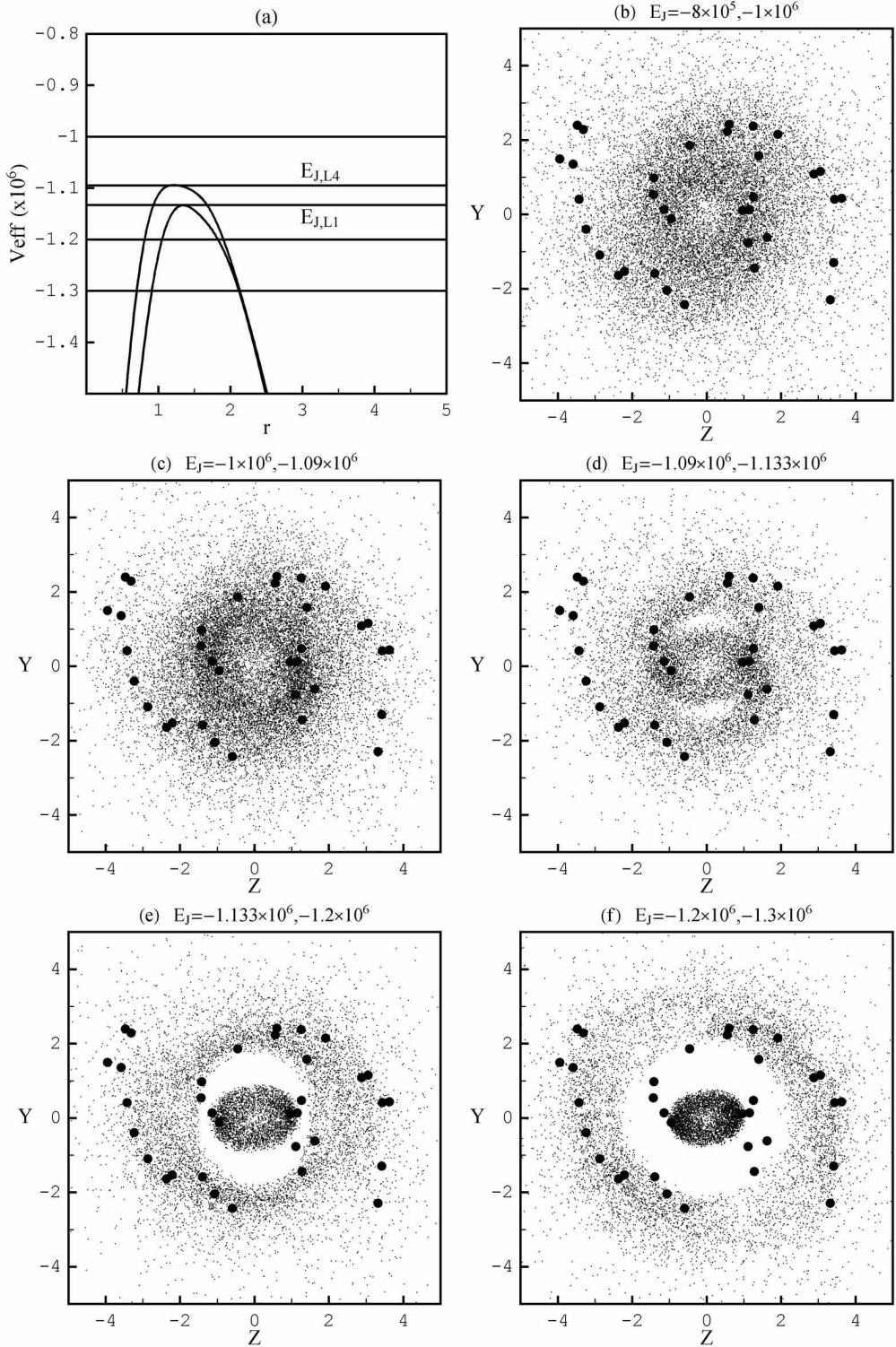}
\caption{(a) Radial profiles of the effective potential $V_{eff}(R)=
V(R,\theta_{fixed})+\Omega^2R^2/2$ along the directions
$\theta_{fixed}$ pointing to $L_1$ (curve with lower maximum) and
$L_4$ (curve with upper maximum). The five horizontal lines denote
the levels of the Jacobi constant to which panels (b) through (f)
refer. These panels show the distribution on the disk plane of the
particles with Jacobi constants belonging to the intervals (b)
$-8\times 10^5\geq E_J\geq -10^6$, (c) $-10^6\geq E_J\geq
-1.09\times 10^6=E_{J,L4}$, (d) $E_{J,L4}=-1.09\times 10^6> E_J\geq
-1.133\times 10^6=E_{J,L1}$, (e) $E_{J,L_1}=-1.133\times 10^6> E_J
\geq-1.2\times 10^6$, and (f) $-1.2\times 10^6> E_J\geq -1.3\times
10^6$. The dots are the same as in Fig.1a.} \label{}
\end{figure}
\begin{figure}
\centering
\includegraphics[scale=1]{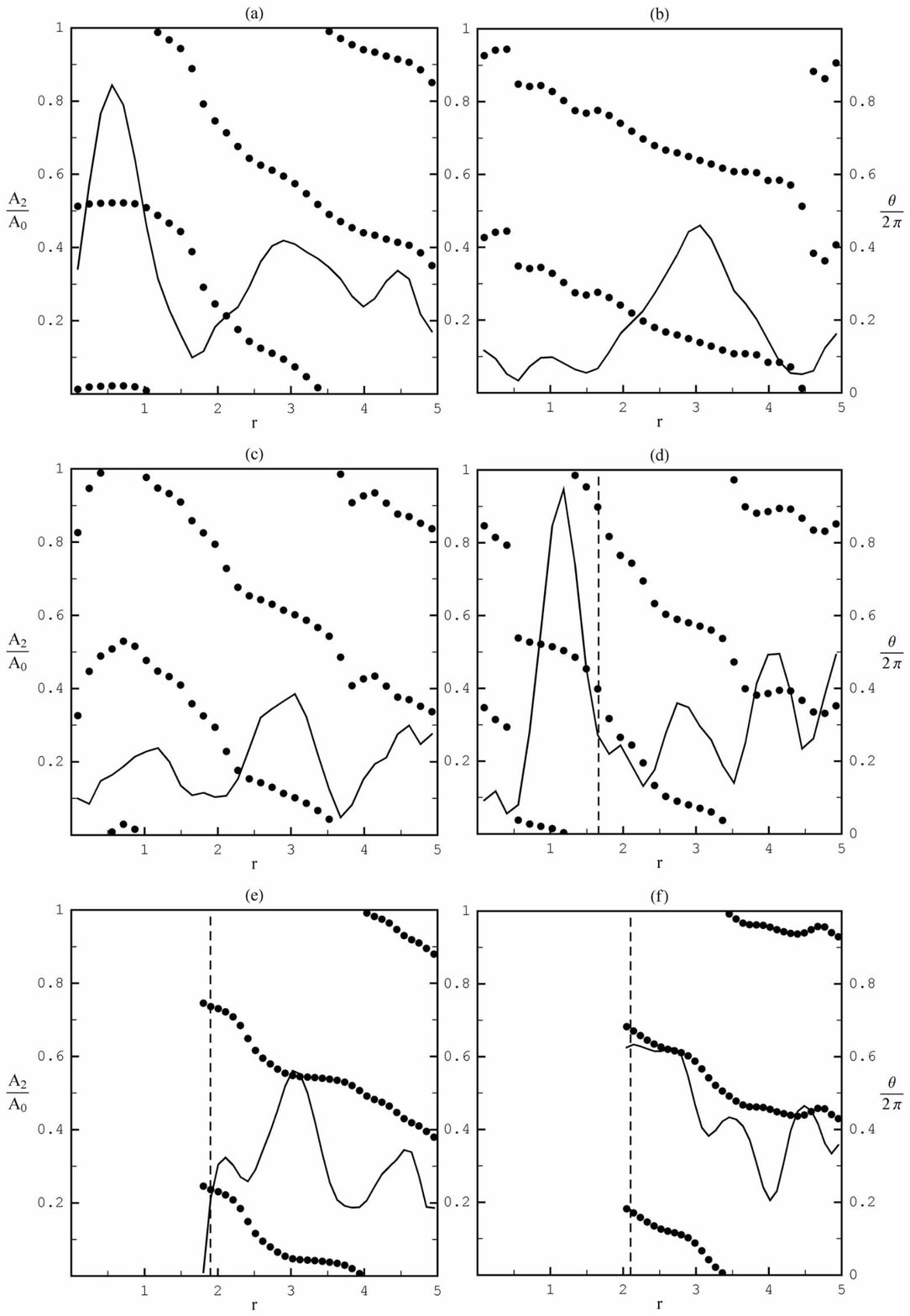}
\caption{ The ratio $A_2/A_0$ (solid lines) of the Fourier
amplitudes $m=2$ vs. axisymmetric of the surface density
$\Sigma(r,\theta)=\Delta N/(r\Delta r\Delta\theta)$, as a function
of $r$, for the particles of (a) the whole simulation, and (b) to
(f) the bins of Figs. 3b to 3f respectively. The black dots show the
values of $\theta_{max}/2\pi$, as a function of $r$, where
$\theta_{max}$ is an angle of maximum of the $m=2$ component for
given $r$. The vertical dashed lines show the limits of $r$ beyond
which annuli of central radius $r$ do not intersect the zero
velocity curves of the effective potential for the lower values of
the Jacobi constant indicated in the panels of Fig.3d to 3f
respectively. In panels e and f the Fourier analysis refers only to
the particles restricted to move outside corotation by the zero
velocity curves. } \label{}
\end{figure}

Figures 3b to 3f show the distribution of the  N-Body particles in
the configuration space for five different bins of the energy
(Jacobi constant) centered at the level values shown in Fig.3a (the
radial profiles of the effective potential $V+\Omega^2R^2/2$ along
the directions crossing $L_1$ and $L_4$ are superposed). The first
bin $=-8\times 10^5\geq E_J\geq -10^6$ (Fig.3b) corresponds to
particles with Jacobi constants well above both
$E_{J,L_1}=-1.133\times 10^6$ and $E_{J,L4}=-1.09\times 10^6$. In
the next bin (Fig.3c) the particles have Jacobi constants just above
$E_{J,L4}$, i.e., $-10^6\geq E_J\geq E_{J,L4} $. In Fig.3d the
particles have Jacobi constants between $E_{J,L4}$ and $E_{J,L1}$.
The remaining two bins (Figs.3e,f) show the particles with Jacobi
constants just below $E_{J,L1}$ ($E_{J,L_1}\geq E_J >-1.2\times
10^6$, Fig.3e) and well below it ($-1.2\times 10^6\geq E_J\geq
-1.3\times 10^6$, Fig.3f) respectively. These energy bins cover an
appreciable range of energies of the histogram of Fig.1a. The thick
dots in panels 3b to 3f are the same as in Fig.1c. The following are
observed:

a) The spiral structure is developed mostly by the particles of Fig.3d,
i.e., in the energy bin $E_{J,L1}\leq E_J\leq E_{J,L4}$, i.e., in the
corotation region.

b) Particles with $E_J<E_{J,L1}$ (Figs.3e,f) contribute to extensions
of the spiral pattern well beyond corotation.

c) Particles with $E_J>E_{J,L_4}$ partly support the spiral pattern
close to $L_{1,2}$ (Fig.3b,c), and partly contribute to the
axisymmetric background. It should be noticed that the distribution
of Fig.1d is obtained by counting all the particles which are inside
bins $r\Delta r\Delta\theta$ at which the projected surface density
presents local maxima. Thus, this distribution contains particles
belonging to the spiral arms as well as to the axisymmetric
background and it is actually impossible to distinguish these two
components by the N-Body data.

A quantitative estimate of the importance of the spiral arms in the
various bins of Figure 3 is obtained in Figure 4, which shows the
profiles of the ratio $A_2/A_0$ (amplitude of the $m=2$ mode over
the axisymmetric background), as a function of the radial distance
$r$, from the Fourier analysis of the projected surface density
$\Sigma(r,\theta)=\Delta N /(r\Delta r\Delta\theta)$ for the total
N-Body simulation and for the separate bins of Figs.3b to 3f. In
each panel of Fig.4, the left vertical axis yields values of the
ratio $A_2/A_0$, and the dependence of this ratio on $r$ is shown by
a solid curve. On the other hand, the right vertical axis in the
same panels shows values of the azimuth $\theta$ normalized over
$2\pi$, and the black dots denote the angles of the maxima of the
$m=2$ mode, again as a function of $r$.

In Fig.4a, these quantities are shown for the surface density of Fig.1a,
i.e., for the whole N-body experiment. The initial rise of the profile
$A_2/A_0$ up to about $r=0.8$ corresponds to the bar $m=2$ amplitude,
which reaches a peak value 0.8. After that, the value of $A_2/A_0$
declines to a minimum $A_2/A_0\simeq 0.1$ at $r=1.6$, but then it
starts rising again, reaching a second maximum $A_2/A_0\simeq 0.4$
at $r=3$. In the whole region from $r=1.5$ to $r=4$ the angles of
$m=2$ maxima $\theta_{max}(r)$ depend monotonically on $r$, and
such a dependence defines spiral arms. After $r=4$ there is a second
local maximum of $A_2/A_0$ which roughly corresponds to the outer
spiral arms described in subsection 2.1.

The remaining panels (Figs.4b to 4f) show the same quantities for
the particles in the corresponding panels of Fig.3. The main
observation here is that the central bin $E_{J,L1}\leq E_J\leq
E_{J,L4}$ (Fig.3d) contributes mostly to the $A_2/A_0$ amplitude at
radii near corotation, i.e., immediately after the bar ($1\leq r\leq
2$, Fig.4d). In particular, very close to the unstable Lagrangian
points (at $r=1.3$) the $A_2/A_0$ ratio approaches unity, and then
it falls rather abruptly to a value fluctuating around
$A_2/A_0=0.3$. On the other hand, in the remaining panels 4b,c,e,
and f the amplitude $A_2/A_0$ becomes important after $r=2$. It
should be noticed that a high value of $A_2/A_0$ (like 0.6 in
Fig.4f) only means that the $m=2$ perturbation is relatively more
important with respect to the axisymmetric background of the
particles in the {\it same} bin (which are, however, fewer than the
particles in the central bin $E_{J,L1}\leq E_J\leq E_{J,L4}$, see
the histogram of Fig.1d). Furthermore, in the bins of Figs.4d,e,f,
which correspond to values of the energy $E_J<E_{J,L4}$, we must
take care of the fact that the motion of the particles is restricted
either inside or outside corotation by the zero velocity curves
(zvc) of the effective potential. Now, because of the presence of
the bar, these curves are elliptically distorted with respect to
perfect circles. Thus, when one makes the Fourier analysis of the
surface density function within different annuli of width $\Delta
r$, if an annulus intersects an inner or an outer limiting zvc at
four bins ($\theta_i\pm\Delta\theta/2$) around respective values
$\theta_i$, $i=1,2,3,4$, there are two symmetric intervals of the
azimuth $\theta$ within which the annulus contains no bodies. This
effect appears as an enhanced amplitude of the m=2 mode in the
Fourier spectrum of $\Delta N(\theta)$ for such an annulus, with the
m=2 maxima being either aligned or at right angles with the bar's
major axis. Thus, this effect is easily distinguished from a real
m=2 mode due to spiral arms. The vertical dashed lines in the panels
of Fig.4d,e,f mark the minimum values of $r$ beyond which annuli of
central radius $r$ do not intersect the zero velocity curves of the
effective potential for the lower values of the Jacobi constant
indicated in the panels of Fig.3d to 3f respectively. In all cases
we find significant $m=2$ amplitudes implying that the spiral arms
are present in all the bins.

\begin{figure}
\centering
\includegraphics[scale=1]{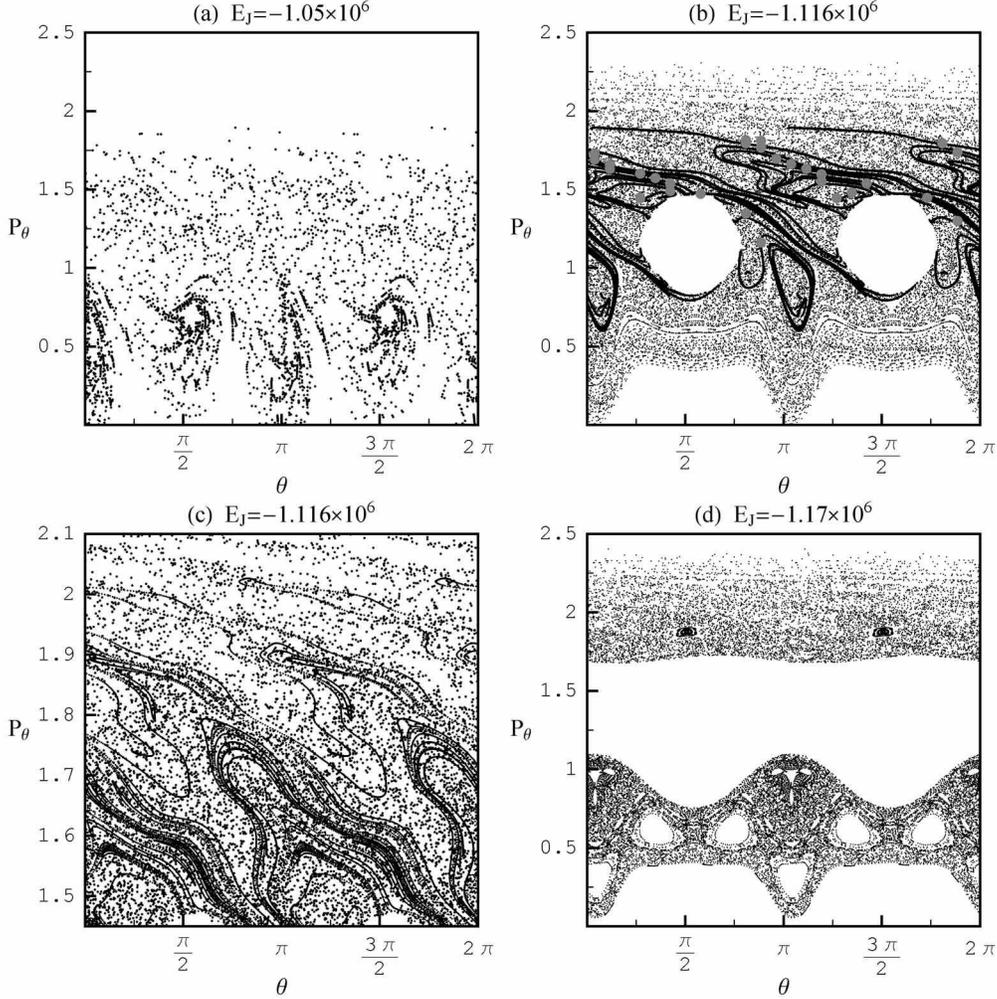}
\caption{ The phase portrait (surface of section
$(\theta,p_\theta)$, with $p_r=\dot{r}=0$ and $\dot{p}_r<0$ ) when
(a) $E_J=-1.05\times 10^6$, (b and c) $E_J=-1.116\times 10^6$, and
(d) $E_J=-1.17\times 10^6$. (c) is a magnification of (b) in the
range of angular momenta $1.45\leq p_\theta \leq 2.1$. The manifolds
in (b) are produced by the 7-th iterates of 10000 initial conditions
along a small segment of length $DS=10^{-4}$ in the ustable
eigendirection of the monodromy matrix at $PL_1$. In (c) the 8-th
iterates of the same manifolds are plotted. The gray thick dots in
(b) correspond to the projection of the locus of maxima of the
spiral pattern (black dots of Fig.1a) on the surface of section
corresponding to a constant energy $E_J=-1.116\times 10^6$. }
\label{}
\end{figure}

Figure 5 shows the phase plots (surfaces of section $(\theta, p_\theta)$) for
three values of the Jacobi constant, namely $E_J=-1.05\times 10^6$ (Fig.5a,
$E_J>E_{J,L4}$), $E_J=-1.116\times 10^6$ (Fig.5b,c $E_{J,L1}<E_J<E_{J,L4}$,
and $E_J=-1.17\times 10^6$ (Fig.5d, $E_J<E_{J,L1}$). These plots are obtained
by integrating the orbits of many test particles with initial conditions
covering the entire plotted domain.

When $E_J>E_{J,L4}$ (Fig.5a) the phase space is open to escapes and,
indeed, we find that for most initial conditions the orbits escape
quickly, typically after less than 10 iterates (= 10 radial
periods). Our numerical criterion for escapes is when an orbit
crosses the radius $r=10r_{corotation}$. In fact, transport from
inside to outside corotation (and vice versa) is energetically
permissible at this value of the Jacobi constant, and we find
numerically that the transport is fast even for orbits with very low
initial values of $p_\theta$, which are placed initially inside the
bar. Nevertheless, some chaotic orbits are also observed in Fig.5a
which are `sticky' to a chain of small islands of stability
corresponing to a 4:1 commensurability. These particles remain
trapped for longer times, which are typically of order $10^2$
periods. The patterns formed in the surface of section
$(\theta,p_\theta)$ by these particles have the typical form
encountered in the so-called outer stickiness zones of the islands
of stability in simple symplectic mappings (Efthymiopoulos et al.
1997). These zones are produced by the invariant manifolds of
unstable periodic orbits of high multiplicity which are located near
the main cantori marking the limits of the islands of stability. The
cantori are invariant sets of points which represent the limit of a
sequence of periodic orbits of progressively higher multiplicity
which exist near the last KAM torus around an island of stability. A
cantorus contains gaps through which chaotic transport is possible.
The main sticky zone of any island is inside a cantorus with small
gaps. The stickiness is caused by the limited flux allowed through
the gaps. This is because the manifolds of the unstable periodic
orbits near the cantorus develop a large number of oscilations in a
direction almost parallel to the cantorus, thus creating a partial
barrier to motions transversely to the cantorus. However, there can
still be appreciable stickiness in a zone outside the cantorus,
caused by the invariant manifolds of unstable periodic orbits
located within that zone (Contopoulos and Harsoula 2007).

When $E_J$ is taken inside the interval $E_{J,L1}<E_J<E_{J,L4}$
(Fig.5b), the picture of the phase space portrait changes
dramatically. Transport from inside to outside corotation is still
energetically permissible within this interval of values of the
Jacobi constant. Furthermore, the phase space is still open to
escapes. However, as observed in Fig.5b, escapes are now much more
difficult, and there is an inner domain (for $p_\theta<2$)which is
almost uniformly covered by the iterates of chaotic orbits. This
domain surrounds two roughly circular domains which are
energetically prohibited to the motion. The inner limit of the large
chaotic domain is defined by a rotational KAM curve, at values of
$p_\theta$ close to and below $p_\theta=0.4$. The regular orbits
below this curve correspond to quasi-periodic motions inside the
bar. On the other hand, for high values of $p_\theta$
($p_\theta>2$), the orbits in the chaotic domain of Fig.5b exhibit,
again, a sticky behavior. Plotting the unstable invariant manifolds
of the $PL_1$ and $PL_2$ orbits in the same figure shows that the
stickiness zone is essentially defined by the outer parts of these
manifolds. A magnification in this region (Fig.5c) clearly shows
that the stickiness zone is structured by the invariant manifolds,
i.e., the stickiness is more pronounced in those domains which are
more densely covered by different parts of the invariant manifolds.
A detailed study of this phenomenon, namely the structuring of the
stickiness zone by invariant manifolds, was made in a simple
dynamical system by Contopoulos and Harsoula (2007).

Now, the gray thick dots in Fig.5b yield the projection on the plane $(\theta,p_\theta)$
of the locus of maxima of the spiral pattern of Fig. 1a. This is found by solving
Eq.(\ref{hamgen}) for $p_\theta$, setting $H=E_J$, $p_r=0$, and $(r,\theta)$
determined by the position of each of the thick dots of Fig.1a. Clearly,
in Fig.5b the maxima of the spiral pattern are also in the same domain as
marked by the invariant manifolds. We conclude that the main action of the
invariant manifolds in this domain is to create stickiness, i.e., to prevent
many stars in chaotic orbits from escaping away from corotation. Furthermore,
during their sticky phase, the chaotic orbits remain in the close neighborhood
of the invariant manifolds, thus contributing to the spiral pattern.

Finally, Fig.5d shows the picture of the phase space when
$E_J=-1.17\times 10^6$, smaller than $E_{J,L1}=-1.133\times 10^6$.
In this case there are two separated domains, lower domain (inside
corotation), and upper domain (outside corotation), in which the
motion is energetically permissible. Communication between the two
domains is not allowed. As shown in Fig.5d, the phase space
structure in the inner domain (for small $p_\theta$) is of `mixed'
type, i.e., there are  some prominent islands of stability
corresponding to either low or high order commensurabilities and
also chaotic zones surrounding the islands of stability. On the
other hand, the upper domain (high values of $p_\theta$) is again
open to escapes, and an outer stickiness zone can be distinguished.
As shown in the next section, this stickiness can no longer be
associated with the invariant manifolds of the $PL_1$ family (which
do not even exist at this value of the Jacobi constant) but it is
rather caused by the invariant manifolds of families of other
unstable periodic orbits with apocenters extending well beyond
corotation. We also show below that these are linked to outer
extensions of the spiral arms as in Figs.3e,f.

\begin{figure}
\centering
\includegraphics[scale=1.2]{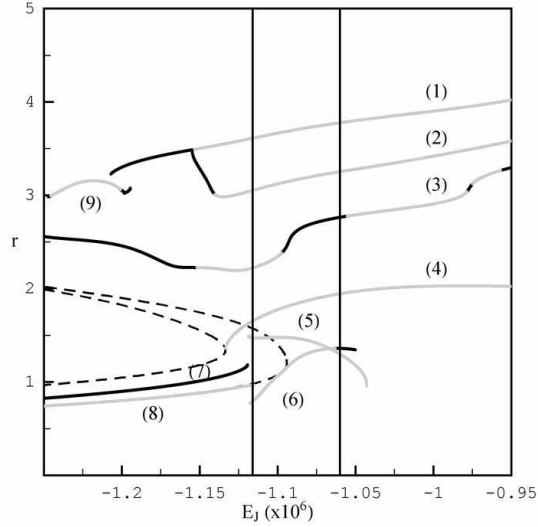}
\caption{ Characteristic curves of nine different families of
periodic orbits: (1) -1:1, (2) -1:1 of multiplicity two (bifurcating
from the -1:1 family), (3) -2:1, (4) $PL_1$, (5) -4:1, (6) 4:1, (7)
and (8) 3:1, (9) -1:1. The ordinate yields the radial distance $r$
at one of the apocenters of the respective periodic orbit as a
function of the Jacobi constant $E_J$. An orbit is stable at black
segments and unstable at gray segments of its characteristic curve.
The commensurabilities of the periodic orbits refer to the number of
radial oscillations per azimuthal period. A minus sign indicates
retrograde azimuthal motion. The profiles of the effective potential
through $L_1$ (inner) and $L_4$ (outer) (same as in Fig.3a) are
overplotted (dashed curves). The left and right vertical lines at
$E_J=-1.116\times 10^6$ and $E_J=-1.06\times 10^6$ pass through
seven and four unstable families respectively. The manifolds of
these families are shown in figures 7a and 10a,b respectively. }
\label{}
\end{figure}

\section{Unstable periodic orbits and their invariant manifolds}

Fig.6 shows the characteristic curves of nine different families of
periodic orbits with apocenters close to or beyond the corotation
region. The characteristic curves correspond to monoparametric
functions $r(E_J)$, $\theta(E_J)$ and $p_\theta(E_J)$ yielding the
apocentric radius, azimuthal angle and angular momentum for which an
orbit with initial conditions $r,\theta,p_\theta,\dot{r}=0$, and
Jacobi constant $E_J$ is periodic. Fig.6 shows only the projection
$r(E_J)$ of each characteristic curve. In fact, contrary to the
usual plots of characteristic curves (see Contopoulos and Grosb{\o}l
1989 for a review of such curves in barred galaxies), the projection
$r(E_J)$ shown in Fig.6 does not suffice to determine the initial
conditions of a periodic orbit, because there is no preferential
axis of symmetry in which the apocenters lie. However, these plots
are indicative of the radial distances reached by each periodic orbit
as a function of the Jacobi constant. The commensurabilities of the
periodic orbits correspond to the ratio of the number of radial
oscilations (apocentric passages) of an orbit per azimuthal period.
Only low order commensurabilities are considered. Finally, the black
and gray parts correspond to intervals of value of the Jacobi constant
at which the corresponding periodic orbit is stable or unstable
respectively.

Some relevant remarks regarding Fig.6 are:

a) For most values of the Jacobi constant beyond $E_{J,L1}$ there are more
than one low order periodic orbits which are unstable. The vertical line
at $E_J=-1.116\times 10^6$ shows an example of a value of $E_J$ at which
the orbits of seven out of the nine considered families are unstable.
The invariant manifolds generated by such orbits are studied in detail
below.

b) After an initial rise to apocentric values $r_a\simeq 1.3r_{corotation}$,
the characteristic curve of the $PL_1$ family extends almost horizontally
to high values of the Jacobi constant. This implies that the apocenters
of the $PL_1$ family remain always relatively close to the corotation region,
and this family plays significant dynamical role for values of $E_J$ well
above $E_{J,L1}$.

c) For most considered values of $E_J$ in Fig.6 we can find at least
one unstable periodic with small absolute value of the H\'{e}non
stability $|b|$ index. Typically, the dependence of the stability
index of one family on $E_J$ shows intervals at which there is an
abrupt rise of $|b|$ to very high values (of order $10^3$ to
$10^4$), followed by other intervals at which $|b|$ falls to
relatively small values. The latter intervals are close to values of
$E_J$ at which there are transitions from stability to instability
(at $|b|=1$). We find numerically that, for almost any value of
$E_J$ in the considered interval, there is at least one unstable
periodic orbit which has a relatively small stability index (below
$|b|<3$). This remark is important, because the time of stickiness
determined by the invariant manifolds of an unstable periodic orbit
depends crucially on the eigenvalue (or stability index) of the
orbit. Namely, the stickiness is appreciable only when the stability
index is close to unity. Thus, there are periodic orbits at almost
any value of the Jacobi constant fulfiling this condition.
\begin{figure}
\centering
\includegraphics[scale=1.1]{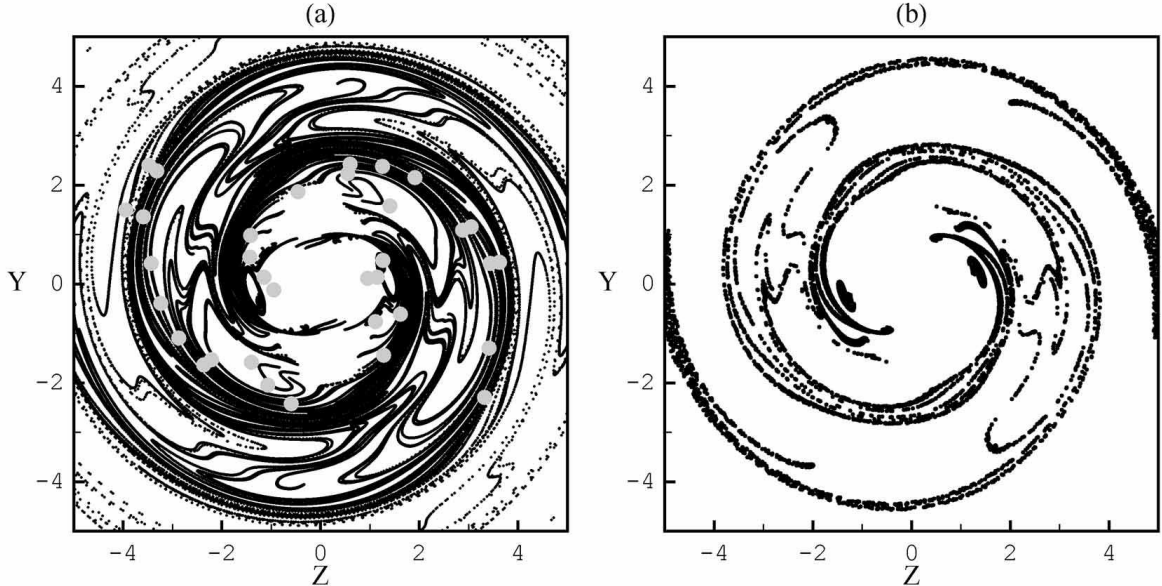}
\caption{ (a) The `coalescence' of invariant manifolds in
configuration space for seven different unstable periodic orbits at
$E_J=-1.116\times 10^6$. The orbits are the unstable orbits crossed
by the left vertical line of Fig.6, namely (1), (2), (3), (4), (5),
(6), and (8). (b) A `response model' with 10000 test particles. The
plot shows the third apocentric positions of all these particles,
when distributed initially uniformly with respect to the angle
$\theta$ along the circular orbit that exists under the monopole
term of the potential expansion at the Jacobi constant
$E_J=-1.116\times 10^6$. The gray dots are the same as the dots of
figure 1a. } \label{}
\end{figure}
\begin{figure}
\centering
\includegraphics[scale=1.1]{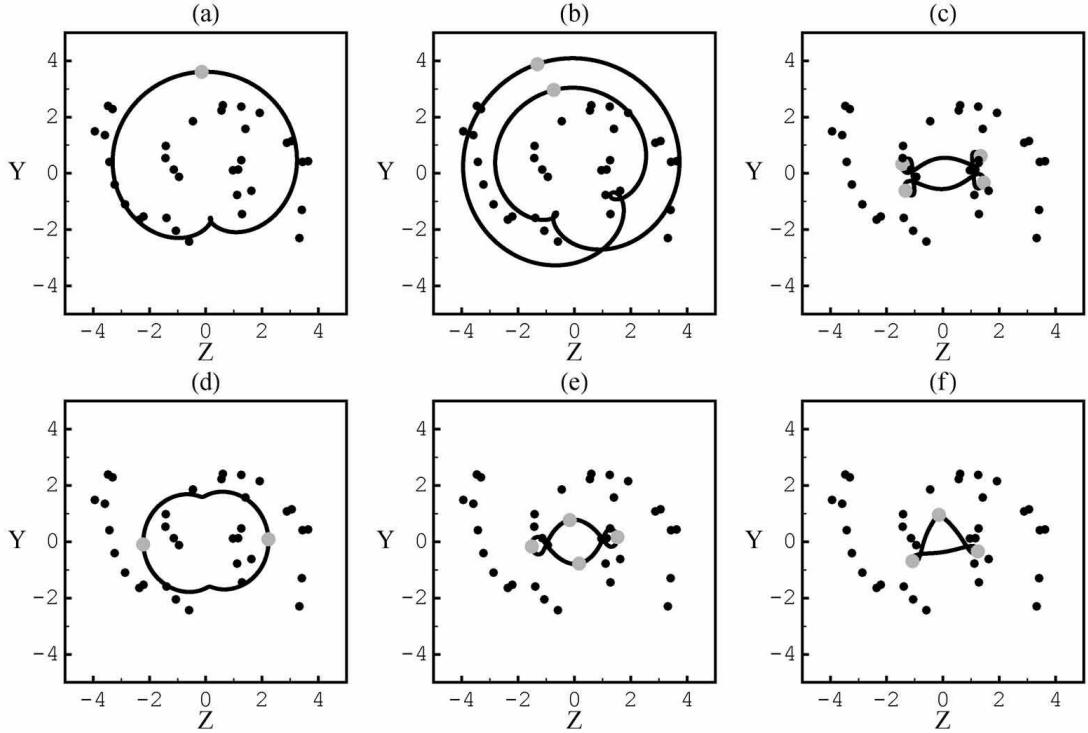}
\caption{The periodic orbits (a)-1:1, (b) -1:1 of multiplicity 2,
(c) -4:1, (d) -2:1, (e) 4:1, and (f) 3:1, at the Jacobi constant
$E_J=-1.116\times 10^6$. The gray dots denote the apocenters of the
orbits. The black dots are the same as in figure 1a.
 } \label{}
\end{figure}
\begin{figure}
\centering
\includegraphics[scale=1.1]{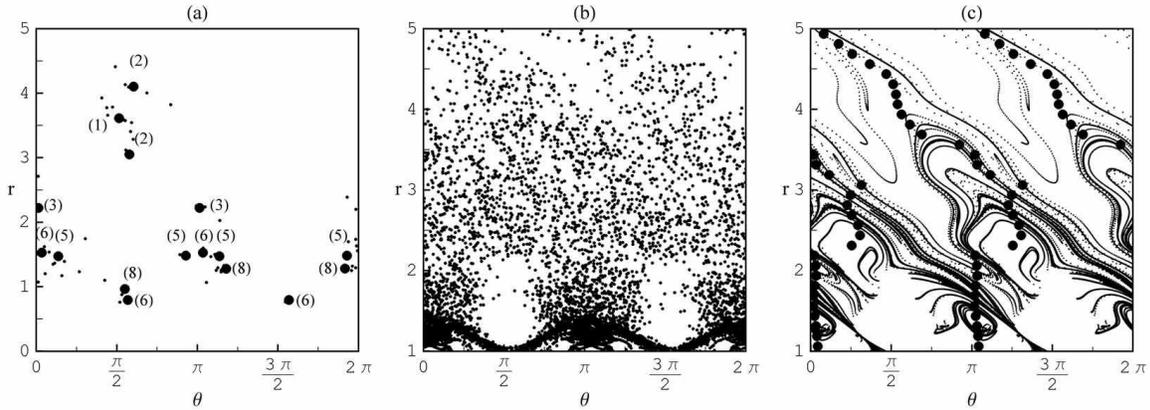}
\caption{(a) The points of all the N-Body particles within the
energy bin $E_{J,L1}\leq E_J\leq E_{J,L4}$ which are in the
immediate phase space vicinity of the periodic orbits of Fig.8, when
the particles' orbits are brought to their nearest consequent with
the surface of section. The particles are located according to the
criterion $d_{min,p}\leq 0.04$. The bold circles denote the position
of the fixed points of the same periodic orbits on the surface of
section. There are 13 nearest particles identified for orbit (1), 6
for orbit (2), 8 for orbit (3), 14 for orbit (5), 13 for orbit (6),
and 18 for orbit (8). (b) Successive consequents of the previous
particles (up to 100 iterations) on the surface of section. (c)
Comparison of the maxima of the density produced by the points of
(b) with the coalescence of the invariant manifolds at
$E_J=-1.116\times 10^6$ (the manifolds of PL1 and -4:1 families are
plotted).} \label{}
\end{figure}

Figure 7 shows now the main result. The unstable manifolds of {\it
seven} different families of periodic orbits, which, for
$E_J=-1.116\times 10^6$ are all unstable, are superposed in the same
plot of the configuration space (Fig.7a). We immediately recognize
that the unstable manifolds of all the families contribute to the
formation of the same spiral pattern. In fact, given that the
unstable manifold of one periodic orbit {\it cannot intersect} the
unstable manifolds of other periodic orbits (see, e.g., Contopoulos
2004, pp.144-157), the manifolds emanating from two different
periodic orbits describe nearly parallel paths in the configuration
space. The so-created set of all the manifolds generates a pattern,
that we call the `coalescence' of invariant manifolds. In our case,
the shape generated by the coalescence of all the manifolds follows
closely that determined by just one manifold, e.g., the manifold of
the $PL_1$ orbit. We emphasize again that each point in Fig.7a
corresponds to an apocentric passage of a chaotic orbit. We thus see
that, as many chaotic orbits, with very different initial
conditions, describe iterations with apocenters along (or close to)
the invariant manifolds of Fig.7a, these orbits are all supporting
the same spiral pattern beyond the bar. Comparison with the thick
dots of the maxima of the real spiral pattern demonstrates that the
real spiral arms are close to the pattern formed by the coalescence
of the invariant manifolds (small differences in the two patterns
are discussed below).

Patsis (2006) recently reported the results of numerical
calculations in a so-called 'response model' of a barred spiral
galaxy, in which, similarly to our example, the spiral pattern is
found to be supported mostly by chaotic orbits. A response model is
created by taking test particles, placed initially near the circular
orbits that exist under the axisymmetric term of the potential.
Then, it is found that, after a short integration, the test
particles with Jacobi constants close to the value at corotation
settle down to positions producing a pattern which has a very good
agreement with the observed spiral pattern. In order to understand
this behavior, we considered in our model the response of the orbits
of 10000 test particles which are initially uniformly distributed
within a narrow strip $(\theta,p_\theta)\in [0,2\pi)\times
[p_{\theta,0}- \Delta p_\theta/2,p_{\theta,0}+\Delta p_\theta/2]$ of
the surface of section for $E_J=-1.116\times 10^6$ (same as in
Figure 7a), where $p_{\theta,0}=1.56$ is the angular momentum
corresponding to a circular orbit in the monopole potential term at
the same value of the Jacobi constant, and $\Delta p_\theta=0.03$.
Figure 7b shows the projection in the configuration space of the
third Poincar\'{e} iterates (=iterates of the mapping on the surface
of section) of all the 10000 orbits. Clearly, after only three
iterates, the orbits of the test particles, which were initially
placed along the corotation circle, respond in such a way that the
apocenters of all the orbits (Fig.7b) are delineated along a locus
which follows essentially the same pattern as that of the invariant
manifolds of Fig.7a, i.e., a spiral pattern (figure 7b does not
change if the non-axisymmetric part of the potential is introduced
gradually). This behavior is understood by noticing that the
coalescence of invariant manifolds within a connected chaotic domain
creates preferential directions along which the chaotic stretching
takes place, so that, after a few transient iterates, the successive
images of a set of nearby initial conditions are delineated along
these directions (see Voglis et al. (1998) for an example of this
behavior in a simple dynamical system). This is precisely what is
observed in Fig.7b, which essentially demonstrates that the
dynamical response of the orbits beyond the bar is determined by the
coalescence of the invariant manifolds.

In order to estimate the relative importance of the various
families, other than $PL_1$ or $PL_2$, in the production of the
spiral pattern due to the coalesence of their manifolds, the
following calculation was made: the six different periodic orbits
were calculated at the Jacobi constant $E=-1.116\times 10^{-6}$
(Figure 8), which is characteristic of the central energy bin
$E_{J,L1}\leq E_J\leq E_{J,L4}$. Then, the N-body particles of the
same bin were identified, which are in the immediate vicinity of
each of these periodic orbits. The numerical criterion for such an
identification was the following: each periodic orbit can be
considered as a smooth curve in phase space, namely it is given by
parametric functions $r_p(s)$, $\theta_p(s)$, $p_{r,p}(s)$,
$p_{\theta,p}(s)$, where $s$ is a parameter along the periodic orbit
$p$. In numerical calculations we set $s\equiv t'$, where $t'$ is
the time it takes to reach a particular point of the periodic orbit
if $t'=0$ is the time at a given initial condition of the orbit.
Let, now, $(r_i,\theta_i,p_{r,i},p_{\theta,i})$ be the position of
the i-th N-body particle of the bin at the snapshot of the
simulation considered. We then calculate the minimum weighted
distance of the body from the periodic orbit given by:
\begin{equation}\label{distmin}
d_{min,p}=\min\Bigg\{
\sqrt{
\big({r_i-r_p(t')\over R_p}\big)^2
+\big({\theta_i-\theta_p(t')\over 2\pi}\big)^2
+\big({p_{r,i}-p_{r,p}(t')\over P_{r,p}}\big)^2
+\big({p_{\theta,i}-p_{\theta,p}(t')\over P_{\theta,p}}\big)^2}~
:~0\leq t'\leq T_{period}
\Bigg\}
\end{equation}
were $R_p$, $P_{r,p}$ and $P_{\theta,p}$ are weighting factors given by:
\begin{equation}
R_p=\int_{0}^{T_{period}}r_p(t')dt',~~~~~~
P_{r,p}=\int_{0}^{T_{period}}|p_{r,p}(t')|dt',~~~~~~
P_{\theta,p}=\int_{0}^{T_{period}}p_{\theta,p}(t')dt'~~~.
\end{equation}
A particle is considered to belong to the $\epsilon-$neighborhood of
the periodic orbit if $d_{min}<\epsilon$. In the numerical test we
set $\epsilon = 0.04$ and have collected 72 particles in total in
the specific energy bin. This method collects particles near every
periodic orbit which are spread all over the orbit. If, however, the
particles' orbits are integrated for a little until they all reach
the surface of section, then the particles yield a concentration of
points on the surface of section near every fixed point produced by
one periodic orbit. These concentrations for the six considered
families are shown in Figure 9a. If now the same particles are
followed for many time steps, they produce an ensemble of points
which all originate from the particular families, i.e., from initial
conditions which are far from the $PL_1$ or $PL_2$ families. This
ensemble is shown on the surface of section of Fig.9b, and it
clearly produces an $m=2$ pattern with the angles of density maximum
depending on $r$ (black points). Superposed to the coalescence of
the invariant manifolds (Fig.9c), these points follow a direction
which clearly coincides with the preferential directions of the
coalescence of invariant manifolds. The numbers of points found
around the various orbits (given in the caption of Fig.9) are
comparable. This fact indicates that particles which are initially
in the vicinity of very different, and distant in phase space,
families of periodic orbits, contribute nearly equally to the
formation of `sticky' chaotic patterns that follow the coalescence
of the invariant manifolds created by all the families.

\begin{figure}
\centering
\includegraphics[scale=1]{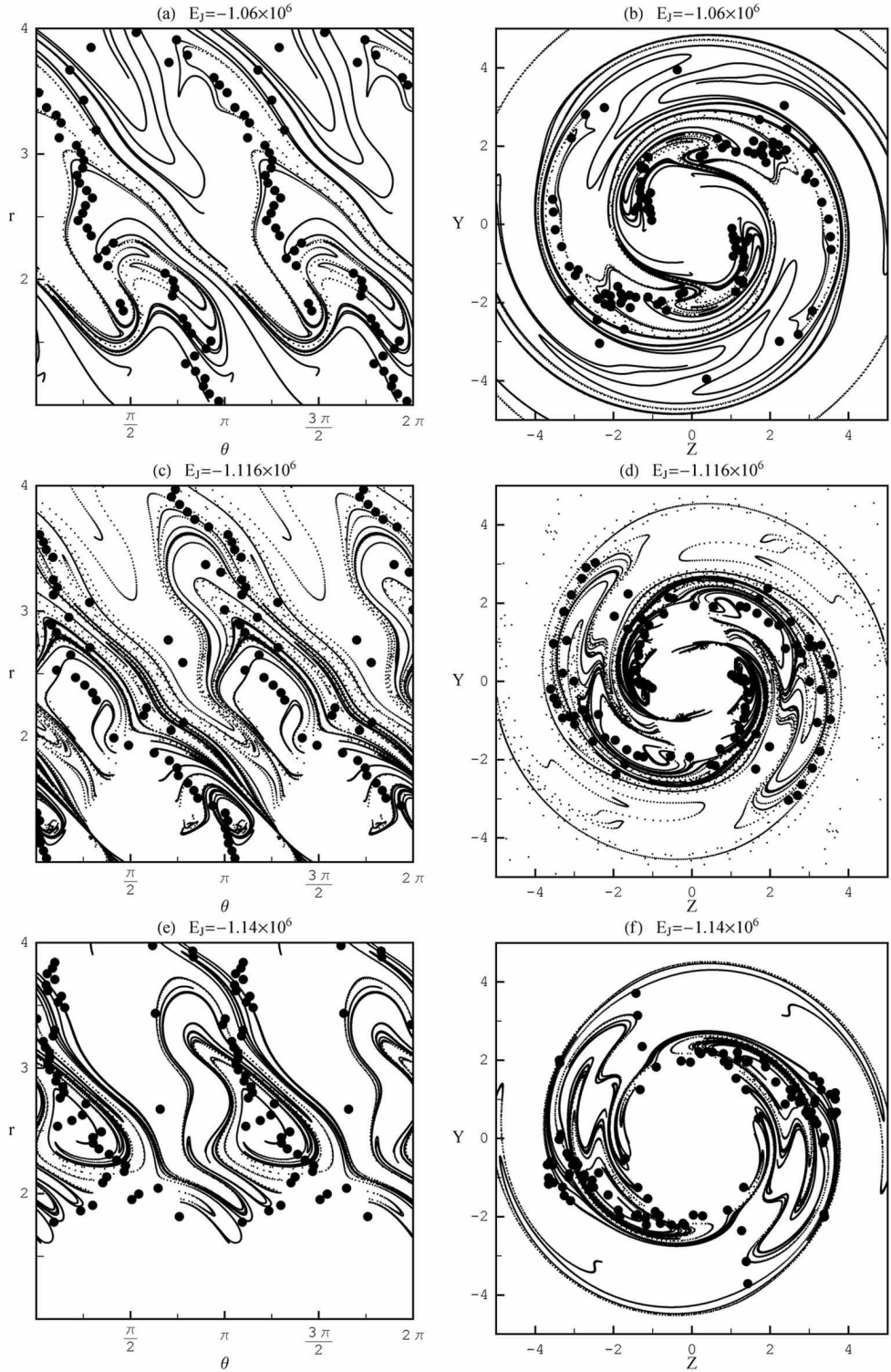}
\caption{ (a,b) Comparison of the loci of the invariant manifolds of
the families (1),(2),(4),(5) at the Jacobi constant $E_J=-1.06\times
10^6$ (right vertical line in Fig.6a), with the maxima of the
density of the particles in the energy bin of Fig.3c, obtained via a
Fourier decomposition keeping up to the m=4 terms (see text). (c,d)
Same for the coalescence of the manifolds at $E_J=-1.116\times 10^6$
and the particles in the energy bin of Fig.3d (the manifolds of the
PL1 and -4:1 families are plotted). (e,f) Same for the manifolds of
the family (3) at the Jacobi constant $E_J=-1.14\times 10^6$ and the
particles in the energy bin of Fig.3e.} \label{}
\end{figure}
As the value of the Jacobi constant increases beyond $E_{J,L1}$, we find a
progressively smaller number of families playing an important role in the coalescence
of invariant manifolds. For example, Figures 10a,b show the superposition of the
unstable manifolds emanating from four different periodic orbits which are unstable
at the value of the Jacobi constant $E_J=-1.06\times 10^6$) (right vertical line in
Fig.7a). At such a value, which is above both $E_{J,L1}$ and $E_{J,L4}$, the
invariant manifolds extend to larger distances away from corotation, but they
also penetrate more deeply inside the bar. The black dots correspond to the
global maxima with respect to $\theta$, for fixed $r$, of the function:
\begin{equation}\label{gaussfour}
D_f(r,\theta)=a_0(r)+a_2(r)\cos 2\theta + b_2(r)\sin 2\theta +
+a_4(r)\cos 4\theta + b_4(r)\sin 4\theta
\end{equation}
where $a_2,b_2,a_4,b_4$ are the coefficients of the Fourier transform, for fixed
$r$, of the function
\begin{equation}\label{gauss}
D(r,\theta)=\sum_{i=1}^{\mbox{N.of particles}}
A\exp\Bigg(-{(r_i-r)^2\over\sigma_r^2}\Bigg)
\exp\Bigg(-{(\theta_i-\theta)^2\over\sigma_\theta^2}\Bigg)
\end{equation}
which represents a Gaussian-smoothed density obtained by the
particles in the same bin of energies as in Fig.3c. The dispersions
$\sigma_r$ and $\sigma_\theta$ are taken equal to the steps
$\sigma_r=\Delta_r$ and $\sigma_\theta=\Delta\theta$ of a $50\times
50$ grid in the square $1\leq r\leq 4$ and $0\leq\theta\leq 2\pi$.
It is necessary to consider the shift of the maxima induced by the
m=4 terms because in some cases such terms reach amplitudes up to
about 1/3 of the m=2 term. We see that the maxima of the particles'
density remain are very close to segments of the manifolds, in
particular those forming bundles of preferential directions. The
maximum phase difference found between the angles of maxima of the
particles' density and the manifolds was 20 degrees, but the typical
difference is of a few degrees. Figures 10c,d show the same
comparison for the coalescence of the manifolds at $E_J=-1.116\times
10^6$, and the particles in the associated energy bin (same as in
Fig.3d). Finally, Figs.10e,f show the same comparison for the
manifolds and particles in the energy bin of Fig.3e, in which the
$PL_1$ family does not exist. This plot shows that despite the
absence of the $PL_1$ family, the manifolds of other families
continue to support the spiral arms. Furthermore, the energetically
allowed outer radial domain is beyond distances $r>1.6$, that is,
the manifolds of the particular families support a part of the
spiral structure that starts well beyond corotation.
\begin{figure}
\centering
\includegraphics[scale=1.2]{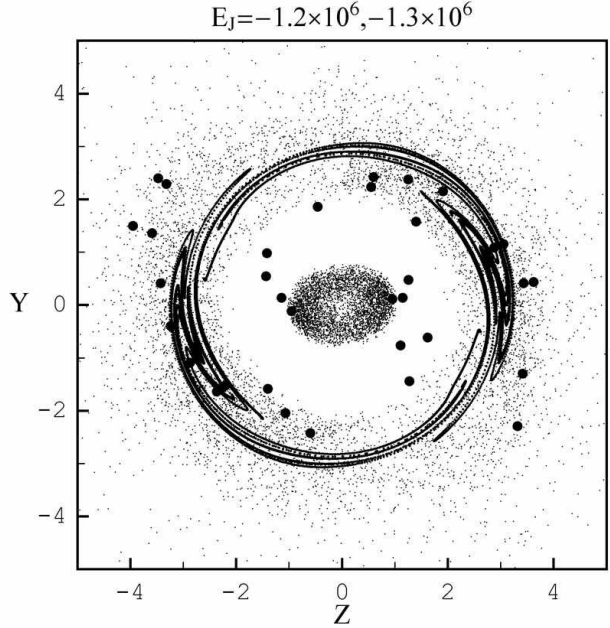}
\caption{Comparison of the loci of the invariant manifolds of the
-1:1 family (curve (9) of Fig.6a) when $E_J=-1.222\times 10^6$, with
the spatial distribution of the N-body particles with Jacobi
constants in the interval $-1.2\times 10^6> E_J\geq -1.3\times
10^6$. } \label{}
\end{figure}
\begin{figure}
\centering
\includegraphics[scale=0.9]{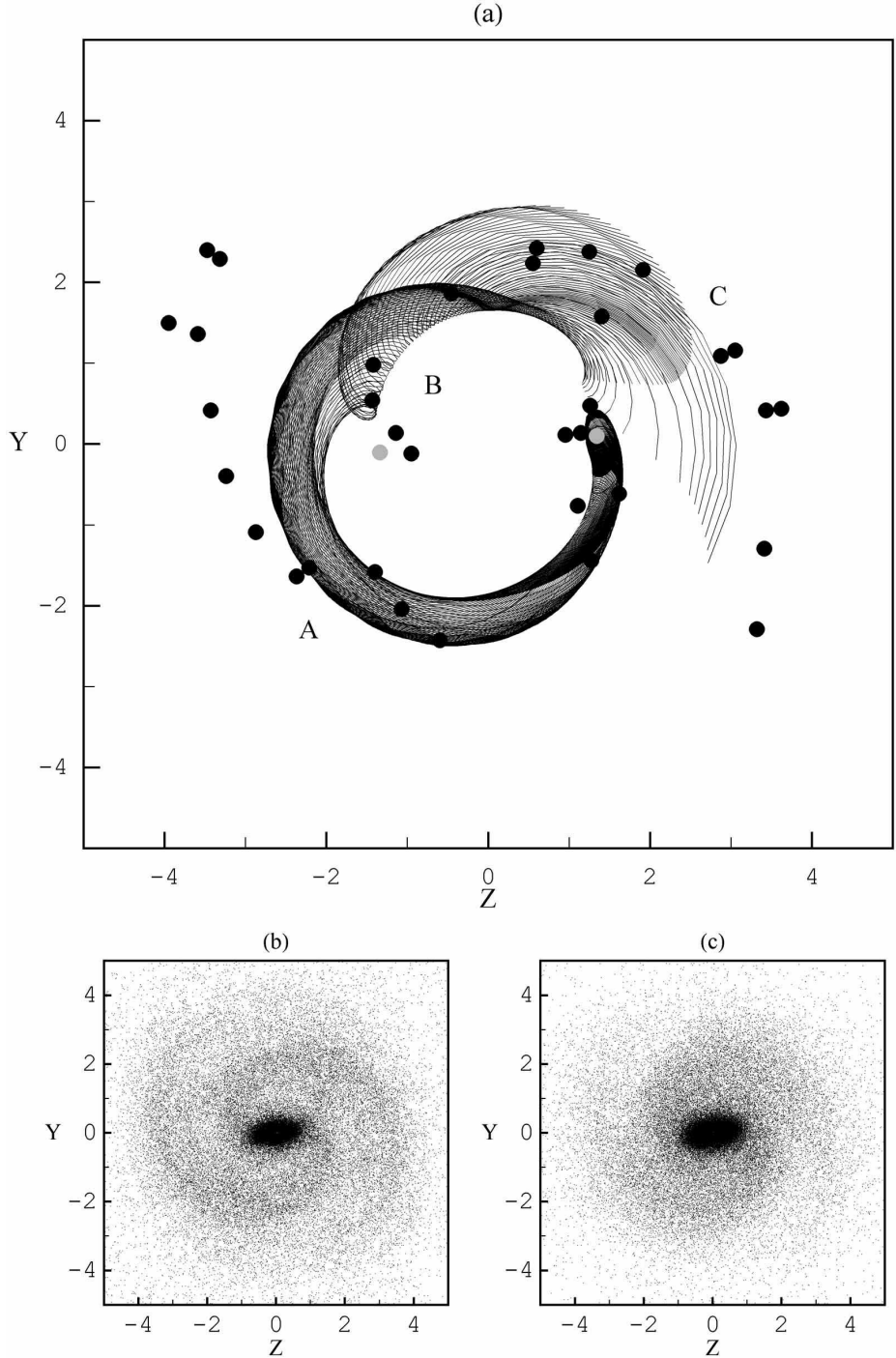}
\caption{(a) Initial segments of the orbits of 20 initial conditions
along the unstable manifold of the $PL_1$ family at $E_J=-1.116\times
10^6$ (b) The projection on the disk plane of the particles with
radial velocities satisfying $|v_r|<0.1\sqrt{2|E_J|}$ at the
snapshot $t=47$. (c) Same as (b) for the particles satisfying
$|v_r|\geq 0.1\sqrt{2|E_J|}$.} \label{}
\end{figure}

As the value of $E_J$ decreases, the bundles of preferential directions become
more and more narrow (Figure 11), and at large distances the patterns generated
by the invariant manifolds become more and more axisymmetric. This tendency of
the manifolds marks the end of the spiral pattern. The manifold of Figure 11
emanates from a -1:1 unstable periodic orbit. This implies that the end of the
spiral pattern in our example is near the outer 1:1 resonance, i.e., beyond the
outer Lindblad resonance.

In all the previous figures, the manifolds are plotted on the surface of section
corresponding to the apocentric positions of the orbits. On the other hand, the
N-body particles are not necessarily close to such positions. A question is,
then, why should the apocentric positions be particularly important in the
determination of the spiral arms. In paper I, this was explained as a consequence
of the fact that the successive apocenters of the chaotic orbits along the manifolds
are well correlated in space and time, their correlation being qualitatively described
by a soliton type equation. However, a quantitative answer can only be based on
numerical evidence, i.e., by considering the behavior in configuration space of
orbits with initial conditions on the invariant manifolds, i.e., asymptotic
to the periodic orbits. As shown in figure~12a, sufficiently close to $L_1$,
segments of orbits asymptotic to a $PL_1$ periodic orbit support the spiral
pattern all along their length, i.e., not only at the apocenters. However,
after an azimuthal turn of about $\theta\simeq 3\pi/4$ from $L_1$, the
asymptotic orbits typically abandon (at point A in figure 12a) the spiral
arm emanating from $L_1$ and form bridges along which material flows to
support either the inner maxima of the $m=2$ pattern( at region B,
pericenters), or the {\it other} arm of the spiral pattern (region C,
apocenters). Orbits having intermediate apocentric passages along the
excursion from one to the other arm create the inner spurs of the invariant
manifolds observed in the interarm region such as in the invariant manifolds
of Figs.7a, or 10.

The importance of the apsidal positions in the support of the
overall m=2 pattern can finally by tested simply by plotting
separately the N-Body particles being close to the apsides at the
given snapshot (figure 12b) from those being far from the apsides
(figure 12c). In these figures the separation of the particles is
based upon a particle's radial velocity $|v_r|$ being smaller or
greated than $0.1\sqrt{2|E_J|}$, where $E_J$ is the particle's
Jacobi constant. Figs.12b,c show that the particles being near their
apsides form a distribution that clearly describes the spiral
maxima, while the particles far from the apsides contribute rather
marginally to the spiral pattern.

All these effects notwithstanding, it should be stated clearly that the theory
developed so far only suggests that the phenomenon of the coalescence of the
invariant manifolds of the unstable periodic orbits near or beyond corotation
constitutes a key concept in order to understand the dynamics of the spiral arms
in barred galaxies. Other phenomena mentioned in section 2, as, for example,
the secular evolution of the bars, or the presence of more than one pattern
speeds, cannot be covered by the presently exposed theory. Thus, a subject
of further study regards incorporating the concept of the coalescence of the
invariant manifolds into a more complete theory for the role of chaos in
the dynamics of the spiral arms in barred galaxies.

\section{Conclusions}

In this paper, which is a continuation of our previous work (paper I, Voglis
et al. 2006), we examine the role played by the invariant manifolds of many
different families of unstable periodic orbits near and beyond corotation
in supporting a spiral pattern beyond the bar of a rotating barred - spiral
galaxy. Our study was based on an N-Body model of such a galaxy. The main
conclusions are the following:

1) One of the main topological properties of the invariant manifolds of
unstable periodic orbits co-existing at the same value of the energy
(Jacobi constant) is that the unstable (stable) manifolds of one family
cannot intersect the unstable (stable) manifolds of any other family.
As a result, the unstable manifolds of all the families are forced to
follow nearly parallel paths in either the phase space or the configuration
space. We call the overall pattern produced by the superposition of the
invariant manifolds of all the different families a `coalescence' of
invariant manifolds.

2) The coalescence of invariant manifolds produces a locus in configuration
space yielding a spiral pattern. Every point on this locus is a local
apocentric position of a chaotic orbit. This locus is invariant under a
Poincar\'{e} mapping of the orbital flow. That is, as the particles on
chaotic orbits are mapped by the orbital flow from one apocentric passage
to the next apocentric passage, the successive apocenters reached by the
particles are all on the same locus determined by the coalescence of
invariant manifolds.

3) The real observed spiral pattern of the system coincides with the
spiral pattern produced by the coalescence of invariant manifolds.
We explain this as a stickiness phenomenon. Namely, while the phase
space is open and the chaotic orbits in general escape within only
a few radial periods to distances far from corotation, the chaotic orbits
with initial conditions close to the invariant manifolds remain trapped
to `sticky' orbits near the manifolds for times of order 100
dynamical periods.

4) Different families of unstable periodic orbits play the dominant
dynamical role in the above phenomena at different values of the
Jacobi constant. The short period unstable family $PL_1$, which
in our previous work (paper I) and in the work of others (Romero-Gomez
et al. 2007) was identified as the main family responsible for the
spiral pattern, is found to play an important role for values of
the Jacobi constant $E_J>E_{J,L1}$, which is however amplified
by the extentions of other families near corotation, e.g., the
3:1 and 4:1 families.

5) The main contribution to the spiral pattern is by the manifolds
of families which are unstable in the energy range $E_{J,L1}\leq
E_J\leq E_{J,L4}$. These manifolds support both the edge of the bar
and the spiral structure close to and outside corotation.

6) The manifolds of families which are unstable for $E_J<E_{J,L1}$ are
responsible for extensions of the spiral arms at distances well beyond
corotation. Beyond the outer Lindblad resonance, the patterns formed
by the invariant manifolds become more and more axisymmetric, and this
tendency marks the end of the spiral arms.

{\bf Acknowledgements:} P. Tsoutsis was supported in part by a
research grant of the Research Committee of the Academy of Athens.
We thank Prof. G. Contopoulos for many suggestions and a careful
reading of the manuscript, as well as an anonymous referee for the
numerous comments that improved the paper considerably. This work
started in collaboration with our teacher and beloved friend N.
Voglis, who passed away on February 9, 2007.

\end{document}